\documentclass[10pt,aps,prd,twocolumn,showpacs,superscriptaddress,nofootinbib,nobibnotes,longbibliography,floatfix]{revtex4-1}

\usepackage[utf8]{inputenc}
\usepackage[T1]{fontenc}
\usepackage{comment}
\usepackage{bm}
\usepackage{mathtools,amsmath,amssymb,amsfonts,mathrsfs,eucal,graphicx,tensor,csquotes,accents,commath,chngcntr,siunitx}
\usepackage[dvipsnames]{xcolor}
\usepackage[unicode]{hyperref}
\hypersetup{colorlinks=true, citecolor=MidnightBlue,
            linkcolor=MidnightBlue, urlcolor=MidnightBlue, linktocpage=true}
\usepackage[normalem]{ulem}
\usepackage{orcidlink}

\DeclareMathAlphabet{\mathpzc}{OT1}{pzc}{m}{it}
\definecolor{darkgreen}{rgb}{0.0, 0.6, 0.0}

\newcommand{\ie}{\textit{i.e.}}

\newcommand{\dd}{\mathrm{d}}

\newcommand{\sst}{\sin^2\!\th}
\newcommand{\cct}{\cos^2\!\th}

\newcommand{\Ord}{\mathcal{O}}

\renewcommand\th{\theta}

\newcommand{\cf}{\varphi}

\begin{document}

\date{\today}

\title{Parameterized Non-circular Deviation from the Kerr Paradigm and Its Observational Signatures: Extreme Mass Ratio Inspirals and Lense-Thirring Effect}

\author{Rajes Ghosh}
\email{rajes.ghosh@iitgn.ac.in}
\affiliation{Indian Institute of Technology, Gandhinagar, Gujarat 382055, India.}

\author{Kabir Chakravarti}
\email{chakravarti@fzu.cz}
\affiliation{CEICO, FZU-Institute of Physics of the Czech Academy of Sciences, Na Slovance 1999/2, 182 21
Prague 8, Czech Republic}

\begin{abstract}
Recent gravitational wave observations and shadow imaging have demonstrated the astonishing consistency of the Kerr paradigm despite all the special symmetries assumed in deriving the Kerr metric. Hence, it is crucial to test the presence of these symmetries in astrophysical scenarios and constraint possible deviations from them, especially in strong field regimes. With this motivation, the present work aims to investigate the theoretical consequences and observational signatures of non-circularity in a unified theory-agnostic manner. For this purpose, we construct a general non-circular metric with a small parameterized deviation from Kerr. This metric preserves the other properties of Kerr, such as stationarity, axisymmetry, asymptotic flatness, and the equatorial reflection symmetry. Apart from the resulting mathematical simplifications, this assumption is crucial to disentangle the consequences of relaxing circularity from other properties. Then, after discussing various novel theoretical consequences, we perform a detailed analysis of extreme mass ratio inspirals and Lense-Thirring precession in the context of this newly constructed metric. Our study clearly shows the promising prospects of detecting and constraining even a slight non-circular deviation from the Kerr paradigm using the future gravitational wave observations by the Laser Interferometer Space Antenna.

\end{abstract}
\maketitle

\section{Introduction}
The unprecedented developments in gravitational wave (GW) astronomy \cite{LIGO1, LIGO2, LIGO3, LIGO4, LIGO5, LIGO6, LIGO7} in the last decade have opened up a new window to understand the workings of gravity, whose classical framework is well-described by Einstein's theory of general relativity (GR). Significant efforts have already been put to test the predictions of GR as well as to explore its limitations, especially in strong field regimes \cite{Will:2014kxa, Psaltis:2008bb, Uzan:2010ri, Yunes:2013dva, Ishak:2018his,  Isi:2019aib, Isi:2020tac, Kramer:2021jcw}. In such extreme gravitational environments, possible deviations from GR are most apparent and likely to be amplified, thereby offering an excellent test bed to identify departures from Einstein's theory that remain obscure in weaker gravitational fields. In this pursuit of detecting and constraining beyond-GR effects, GWs stand out as the most promising tool \cite{Barack:2018yly, Yunes:2013dva} for their ability to encompass the fundamental attributes of gravity across a wide range of length scales.\\

In the quest of venturing beyond Einstein's theory, one usually uses the framework of modified gravity to effectively capture possible deviations from GR. These alternative theories either change the gravitational dynamics by introducing higher curvature terms \cite{Sotiriou:2008rp, DeFelice:2010aj, Salvio:2018crh, Kim:2019sqk, Fernandes:2022zrq, Padmanabhan:2013xyr} (and/or additional fields \cite{Fujii:2003pa, Faraoni:2004pi, Horndeski:1974wa, Kobayashi:2019hrl}) or model the departures using well-motivated/observation-oriented phenomenological means \cite{Gair:2007kr, Johannsen:2011dh, Johannsen:2013szh, Rezzolla:2014mua, Konoplya:2016jvv}. Both of these avenues have been extensively studied in literature and have offered promising results \cite{Faraoni:2004pi, Kobayashi:2019hrl, Ryan:1997hg, Glampedakis:2005cf, Nojiri:2008nt, Barausse:2008xv, Evstafyeva:2022rve, Dey:2022pmv}. However, apart from the dynamical/phenomenological alteration of GR, it is also important to probe various kinematical aspects of gravity that are theory-agnostic and universal in nature. Often a great deal of such kinematical structure follows as a direct consequence of various spacetime symmetries. For example, the planar nature of Schwarzschild geodesics can be traced back to the underlying spherical symmetry. Similarly, in Kerr spacetime, the presence of Carter constant plays a crucial role to understand the nature of timelike/null geodesics \cite{Carter:1968rr}.\\

Apart from various obvious symmetries such as stationarity and axisymmetry, the Kerr metric is also \textit{circular} \cite{Carter:1973, Waldbook}. It is essentially an integrability property that warrants the existence of only a single cross-term ($g_{t\cf}$) in the metric when expressed in terms of the Boyer-Lindquist coordinates. As a consequence, under a simultaneous flip of the direction and axis of rotation, the spacetime metric and all physical quantities remain invariant. Moreover, a black hole (BH) spacetime being circular entails some important properties, such as the constancy of horizon angular velocity \cite{Frolov:1998wf} and validity of the zeroth law of BH mechanics \cite{Heusler}.\\

In vacuum GR, circularity follows as a direct consequence of Einstein's field equations \cite{Waldbook, Kundt:1966}. However, as we venture beyond vacuum GR or in modified gravity, stationary and axisymmetric spacetimes need not be circular. In recent years, a renewed interest has emerged to study both theoretical and observational implications of circularity \cite{Anson:2020trg, Xie:2021bur, Anson:2021yli, Takamori:2021atp, Delaporte:2022acp, Chen:2022ynz, Babichev:2024hjf}. Interestingly, it has been demonstrated under certain conditions that for any effective theory of gravity perturbatively connected to GR, a BH solution remains circular at all orders when expanded in series of coupling constant(s) \cite{Xie:2021bur}. However, when some of these conditions are relaxed, one indeed obtains non-circular BH solutions. One such example that has been well-studied in literature is the so-called DHOST solution \cite{Anson:2020trg, BenAchour:2020fgy, Anson:2021yli}.\\

However, it seems rather impractical and almost impossible to find and study non-circular rotating BH solutions on a theory-by-theory basis, even if certain reasonable assumptions like asymptotic flatness are assumed. Thus, the question still remains whether one can study the theoretical and observational consequences of non-circularity in a theory-agnostic and unified manner. This is what motivates our present work which proposes a parameterized non-circular BH metric slightly deviated from Kerr, as both shadow imaging and GW observations are consistent with the Kerr paradigm. All other properties of the Kerr metric, like stationarity, axisymmetry, asymptotic flatness, and the $\mathbb{Z}_2$ reflection symmetry about the equatorial plane, are kept intact. Besides the obvious mathematical simplifications, this assumption is absolutely necessary to distinguish the consequences of relaxing circularity from other properties. However, as demonstrated in Refs.~\cite{Delaporte:2022acp, Benenti:1979erw, Bezdekova:2022gib, Chen:2023gwm}, geodesic-separability and the presence of Carter constant (originating from a rank-2 Killing tensor) necessarily imply circularity\footnote{We thank Che-Yu Chen and Hsu-Wen Chiang for pointing this out to us and many helpful discussions on similar topic.}. Hence, we construct our beyond-Kerr metric in such a way that the absence of Carter constant is due to the underlying non-circularity of the metric.\\

We then present a rigorous discussion of various mathematical properties of this newly constructed BH metric, such as the location of the event horizon and ergoregion, existence of light ring(s), and the presence of innermost stable circular orbit (ISCO). Apart from these theoretical studies, any modification of Kerr paradigm should also be confronted with observations, such as GW observations by LIGO-Virgo-KAGRA (LVK) collaboration \cite{LIGO1, LIGO2, LIGO3, LIGO4, LIGO5, LIGO6, LIGO7}, and BH shadow imaging by the Event Horizon Telescope \cite{EHT1, EHT2, EHT3, EHT4, EHT5, EHT6}. However, since the non-circular deviation beyond Kerr is assumed to be small, both of these observations might not be enough for constraining the deviation parameters. In particular, the ambiguities in modelling the environment for shadow observations, and the present lack of GW signals which are both long and loud are potential roadblocks that dilute the effects of non-circularity. For this reason, the framework of \textit{extreme mass ratio inspirals} (EMRIs) \cite{Ryan:1995zm, Ryan:1995xi, Drasco:2005kz, Hinderer:2008dm, LISA:2022kgy, LISA:2022yao}, in which a solar-mass secondary object orbits around a supermassive BH (SMBH), seems excellently suited to the task. EMRIs inspiral in the in-band orbit for months, providing us with a faithful map of the spacetime outside the central SMBH. Therefore, even a small deviation from the Kerr paradigm, such as due to non-circularity, can accumulate to detectable levels in observations by Laser Interferometer Space Antenna (LISA) \cite{LISA:2022kgy, LISA:2022yao, Barack:2003fp, Gair:2011ym, Seoane:2017, Babak:2017tow, Colpi:2024xhw} (see also Ref.~\cite{Cardenas-Avendano:2024mqp, Rahman:2023sof, Rahman:2022fay, Rahman:2021eay, AbhishekChowdhuri:2023gvu, Kumar:2024utz}). This motivates us to study the GW emission and the corresponding orbital shrinkage for an equatorial EMRI system, where the spacetime outside the primary is modelled by our non-circular metric given in Eqs.~\eqref{g1asymp} and \eqref{fhsymp}.\\

Apart from EMRIs, we shall also consider the \textit{Lense-Thirring} (LT) effect \cite{deSitter:1916zz, Lense:1918, Schiff:1960, Hartle:2009}, whereby a test gyroscope outside a rotating object precesses with a certain frequency $\Omega_{LT}$ due to frame-dragging. In a Kerr spacetime, this LT frequency is well-known in literature \cite{Straumann:2013spu, Chakraborty:2013naa} and its magnitude has the far-field form up to $\mathcal{O}(r^{-4})$,
\begin{equation}\label{LTk}
    \Omega_{LT}^{(0)}(r,\th) \approx \frac{a M \sqrt{3\cos^2\th +1}}{r^3}+\frac{2 a M^2 (\cos^2\th +1)}{r^4 \sqrt{3\cos^2\th +1}}\, . 
\end{equation}
It is worth mentioning that both the LAGEOS experiment \cite{Ciufolini:2004rq} and Gravity Probe B \cite{Everitt:2011hp} have confirmed the consistency of the above expression in earth's gravitational field. In this work, we want to calculate the deviation from $\Omega_{LT}^{(0)}$ due to non-circularity, which will provide a powerful way to constrain various model-parameter(s).\\

Finally, we conclude discussing various future prospects of constraining non-circular deviation parameters using other considerations/observations. Among these, a detailed study of BH quasi-normal modes and tidal response might prove useful.

\section{Kerr spacetime and its symmetries}
To construct a non-circular metric, it is useful to review some important features of the Kerr metric first. The Kerr spacetime represents the unique $4$-dimensional stationary, axisymmetric and asymptotically flat rotating vacuum BH solution in GR \cite{Kerr:1963ud, Heusler, Mazur:2000pn, Robinson:2004, Chrusciel:2012jk}. As a consequence of the celebrated ``no-hair theorem'' \cite{Carter:1971zc, Bekenstein:1971hc, Bekenstein:1972ky}, a complete specification (e.g. all mass and spin multipole moments) of the Kerr spacetime is characterized solely by two parameters, namely the mass $M$ and the spin $a$ of the BH \cite{Hansen:1974zz, Geroch:1970cd}. In Boyer-Lindquist coordinates $x^\mu = (t,r,\th,\cf)$, the line element of the Kerr metric is given by
\begin{equation}
\begin{split}\label{kerr}
    & \dd s^2_{(0)}=-\left(1-\frac{2 M r}{\rho^2}\right)\, \dd t^2 + \frac{\rho^2}{\Delta}\,  \dd r^2\\
    &+\rho^2\, \dd\th^2 +\sst 
    \left[\frac{(r^2+a^2)^2-\Delta a^2 \sst }{\rho^2}\right]\, \dd\cf^2 \\
    &-\frac{4 M r a \sst}{\rho^2}\, \dd t\, \dd\cf,
\end{split}
\end{equation}
where $\rho^2 = r^2+a^2\, \cct$, and $\Delta = r^2+a^2-2 M r$. The event horizon, which is also a Killing horizon, is located at $r_H = M + \sqrt{M^2-a^2}$. It is easy to notice the presence of two Killing isometries $\xi=\partial_t$ and $\chi=\partial_\cf$ of the metric. Moreover, the spacetime is $\mathbb{Z}_2$ symmetric for reflection across the equatorial plane $(\theta = \pi/2)$. As a result, the equatorial Kerr geodesics are planar. \\

Apart from these, the above metric also possesses some other important symmetries. For example, the Kerr spacetime is circular \cite{Carter:1973, Waldbook}, representing the fact that the metric has only one cross-term (the $\dd t\, \dd\cf$ term) and hence, it is invariant under a simultaneous coordinate transformation $\{t \to -t, \cf \to -\cf\}$. In fact, in vacuum GR, circularity is a direct consequence of the field equations and asymptotic flatness \cite{Papa:1966, Carter:1973, Waldbook, Kundt:1966}. Geometrically, it guarantees the existence of a family of $2$-dimensional surfaces, called ``surfaces of transitivity'', to which the Killing vectors are everywhere tangent except on the rotation axis where $\chi$ vanishes \cite{Xie:2021bur}. In other words, circularity is an integrability property of the spacetime that warrants the existence of two meridional coordinates $(r,\th)$ such that $g_{t r}=g_{t \th}=g_{\cf r}=g_{\cf \th}=g_{r \th}=0$.\\

There is also a coordinate-free notion of circularity \cite{Frolov:1998wf, Stephani:2003tm}, which claims that the necessary and sufficient condition for a metric to be circular is (for all values of $r$ and $\th$)
\begin{equation} \label{circ}
\xi \wedge \chi \wedge \dd \xi=\xi \wedge \chi \wedge \dd \chi=0\, .
\end{equation}
In fact, the above conditions can be used to constrain various components of the Ricci tensor \cite{Kundt:1966, Papa:1966, Waldbook, Xie:2021bur}. Then, it is not hard to show that circularity implies the constancy of angular velocity $\Omega_H$ of the event horizon, i.e., $g_{\cf \cf}^2 \Omega^H_{[,\alpha}\, \beta_{\mu \nu]} = 0$, where $\beta_{\mu \nu} = 2\, \xi_{[\mu} \chi_{\nu]}$ is a bivector null on the event horizon \cite{Frolov:1998wf}. However, we must emphasize that the loss of circularity does not necessarily imply that $\Omega_H$ will vary across the event horizon.\\

Besides circularity, the Kerr metric also has an extra symmetry leading to geodesic separability. This symmetry is linked with the existence of the Carter constant \cite{Carter:1968rr, Carter:1973}, originating from a Killing tensor \cite{Walker:1970un, Waldbook}. In particular, using Hamilton-Jacobi method, one can decouple the radial and angular parts of the Kerr geodesics as follows \cite{Carter:1968rr, Carter:1973}:
\begin{equation}
\begin{split}\label{kerrsep}
    &-\Delta\, p_r^2 + \frac{1}{\Delta}\Big[(r^2+a^2)E-a\, L_z \Big]^2-m^2\, r^2\\
    &= p_\th^2+\Big[ \frac{L_z}{\sin \th}-a\, E\, \sin\th\Big]^2+m^2\, a^2\, \cos^2\th = Q\, ,
\end{split}
\end{equation}
where $m = 0$ and $(\text{mass})^2$ for photons and massive particles, respectively. Moreover, $E$, $L_z$, and $Q$ are the conserved energy, z-component of the angular momentum, and the reduced Carter constant of the geodesic particle with $4$-momentum $p_\mu$. It is needless to say that separability plays an important role in understanding the geodesic properties of the Kerr spacetime.\\

Recently, it has been demonstrated that the non-existence of Carter constant gives rise to some non-trivial features in the corresponding GW signatures \cite{Destounis:2021mqv, Destounis:2021rko, Chen:2023gwm}, providing us with a novel way to detect departures from separability. However, no such universal signatures of non-circularity have been identified till date, which is mainly due to the absence of a general non-circular metric. This is what we aim to construct in the following section. 

\section{Construction of a parameterized non-circular metric}
With all the special symmetries, one would think that Kerr metric is only suitable for some ideal scenarios and might fail to represent rotating astrophysical BHs. In contrary, recent shadow imaging \cite{EHT1, EHT2, EHT3, EHT4} and GW observations \cite{LIGO1, LIGO2, LIGO3, LIGO4, LIGO5, LIGO6, LIGO7} have clearly demonstrated that the rotating BHs in the universe are fairly well-described by the Kerr metric. Hence, the consistency of \textit{Kerr paradigm} highly constrains any possible deviation from the Kerr metric, arising due to modified gravity framework or non-trivial environmental effects. Any such deviations are potential smoking gun for detecting new physics in the strong field regimes of gravity, where GR still lacks enough observational supports. This is what motivates the extensive study of various post-Kerr metrics in the literature \cite{Gair:2007kr, Johannsen:2011dh, Johannsen:2013szh, Rezzolla:2014mua, Konoplya:2016jvv}. Most of these metrics are constructed assuming the deviations beyond Kerr paradigm only modifies the metric components without changing the underlying symmetries. However, since several properties of Kerr spacetime follows directly from various symmetries, it is important to study the departures from Kerr paradigm that considers relaxing some of these symmetries.\\

Motivated by the above discussion, we shall now construct a general parameterized non-circular metric that is ``slightly'' deviated from that of Kerr. Since the main goal of this work is to understand the consequence of circularity, we shall also assume the presence of other properties of Kerr, such as stationarity, axisymmetry, asymptotic flatness, and the equatorial reflection symmetry, are kept intact. Apart from the resulting mathematical simplifications, this assumption is crucial to disentangle the consequences of relaxing circularity from other symmetries. With this, we may express a general non-circular, stationary and axisymmetric metric in $4$-dimension as,
\begin{equation}
\begin{split}\label{metric}
    \dd s^2=&\, g_{tt}\, \dd t^2 + g_{rr}\,  \dd r^2
    +g_{\th \th}\, \dd\th^2 +g_{\cf \cf}\, \dd\cf^2
    +2g_{t \cf}\, \dd t\, \dd\cf \\
    &+2g_{r \th}\, \dd r\, \dd \th+2g_{t r}\, \dd t\, \dd r+2g_{t \th}\, \dd t\, \dd\th \\
    &+2g_{r \cf}\, \dd r\, \dd\cf+2g_{\th \cf}\, \dd \th\, \dd\cf.
\end{split}
\end{equation}
Note that all metric components are functions of $(r,\th)$ alone. Moreover, since the metric is assumed to be slightly deviated from that of Kerr, we may write $g_{\mu \nu}(r,\th) = g_{\mu \nu}^{(0)}(r,\th)+\epsilon\, g_{\mu \nu}^{(1)}(r,\th)$ \footnote{The number inside $(..)$ is to distinguish the Kerr metric components from the deviations. For the ease of writing, we shall put $(..)$ either as superscript or subscript.}. Here, $g_{\mu \nu}^{(0)}$ represents the Kerr metric given by Eq.~\eqref{kerr}, whereas $g_{\mu \nu}^{(1)}$ represents the non-circular departure proportional to a small dimensionless deviation parameter $\epsilon$ (with $|\epsilon| \ll 1$). In all subsequent expressions, we shall only keep terms up to the linear order in $\epsilon$. Moreover, since the Kerr metric is circular, we must have $g_{t r}^{(0)}=g_{t \th}^{(0)}=g_{r \cf}^{(0)}=g_{\th \cf}^{(0)}=0$, and the radial coordinate can be chosen in such a way that $g_{r \th}^{(0)} = 0$.\\

Due to the freedom of arbitrary diffeomorphisms, the metric in Eq.~\eqref{metric} can have at most $10-4 = 6$ independent functions. However, further considerations (e.g. asymptotic flatness) can reduce this number below six. Now, since we are interested in performing Hamilton-Jacobi separability, we need the contravariant metric components $g^{\mu \nu} = g^{\mu \nu}_{(0)}+\epsilon\, g^{\mu \nu}_{(1)}$ as given by Appendix-\ref{app:inverse}. 

\subsection{Hamilton-Jacobi Method and Geodesic Separability}

Following a similar analysis for the Kerr case \cite{Carter:1968rr, Carter:1973}, let us now study the timelike and null geodesics of the spacetime given by Eq.~\eqref{metric}. For an affinely parameterized geodesic, we have $g^{\mu \nu}\, p_{\mu}\, p_{\nu} = -m^2$, where $m$ is either $0$ or $(\text{mass})^2$ depending on whether the particle is photon or a massive one with momentum $p_\mu$. Then, in order to study the geodesic separability, it is customary to introduce a Jacobi function $S = (m^2/2)\, \lambda-E\, t+L_z\, \cf+S_r(r)+S_\th(\th)$, where $\lambda$ is an affine parameter and $S_{,\mu}$ is identified with $p_\mu$. Note that, in general, the previous ansatz for $S$ will not be consistent. However, since we want to force geodesic separability on the metric, this ansatz is well-motivated.\\

After some algebraic manipulations, one can show that the equation $g^{\mu \nu}\, p_{\mu}\, p_{\nu} = -m^2$ takes the following form,
\begin{equation}
\begin{split}\label{metricgeo}
    &\left[ \Delta\, S'_r(r)^2 - \frac{1}{\Delta}\Big\{(r^2+a^2)E-a\, L_z \Big\}^2+m^2 r^2 \right]+\\
    &\left[ S'_\th(\th)^2+\Big\{\frac{L_z}{\sin \th}-a E \sin\th\Big\}^2+m^2 a^2 \cos^2\th \right] + \epsilon\, L= 0,
\end{split}
\end{equation}
where the last term is $L(r,\theta) = \rho^2\, g^{\mu \nu}_{(1)}\, S_{,\mu}\, S_{,\nu}$, which will prevent the above equation to decouple into separate radial and angular parts unless $L(r,\th)=F(r,p_r)+H(\th, p_\th)$ for some functions $F$ and $H$.\\

This will, in turn, constrain the functional dependence of various metric components. In particular, using the results in Appendix-\ref{app:inverse}, it is easy to see that the necessary and sufficient conditions for separability up to the linear order in $\epsilon$ are as follows:\\
(i) $g^{r \th}_{(1)} \sim g_{r \th}^{(1)}$ must vanish as it introduces an explicit coupling in the $r\th$-sector.\\
(ii) $\rho^2\, g^{\mu \th}_{(1)}$ (with $\mu \neq r$) has to be a function of $\th$ alone, say $h_{\mu \th}(\th)$ (symmetric).\\
(iii) $\rho^2\, g^{\mu r}_{(1)}$ (with $\mu \neq \th$) has to be a function of $r$ alone, say $\Delta(r)\, f_{\mu r}(r)$ (symmetric). The extra factor of $\Delta$ is to make the subsequent expressions look cleaner.\\
(iv) $\rho^2\, \left(g^{tt}_{(1)}\, E^2 -2\, g^{t \cf}_{(1)}\, E\, L_z+g^{\cf \cf}_{(1)}\, L_z^2\right)$ has to be of the form $f(r)+h(\th)$ for all values of $(E,L_z)$.\\

Now, along with $g^{\mu \nu}$ given in Appendix-\ref{app:inverse}, the above conditions can be inverted to obtain $g_{\mu \nu}^{(1)}$ uniquely. These deviation metric components up to the linear order in $\epsilon$ are given by,
\begin{equation}
\begin{split}\label{g1}
    &g_{tt}^{(1)}= -\frac{1}{\rho^2} \Big[g_{t t}^{(0)2}\, F_{tt}+2\, g_{t \cf}^{(0)}\, g_{tt}^{{(0)}}\, F_{t \cf} + g_{t \cf}^{(0)2}\, F_{\cf \cf}\Big], \\
    &g_{t\cf}^{(1)}= -\frac{1}{\rho^2} \Big[g_{t \cf}^{(0)}\, g_{t t}^{(0)}\, F_{tt}+ g_{t \cf}^{(0)}\, g_{\cf \cf}^{(0)}\, F_{\cf \cf} \\& \hskip 10em + \left( g_{t \cf}^{(0)2}+g_{t t}^{(0)}\, g_{\cf \cf}^{{(0)}}\right)F_{t \cf}\Big], \\
    &g_{\cf \cf}^{(1)}= -\frac{1}{\rho^2} \Big[g_{t \cf}^{(0)2}\, F_{tt}+2\, g_{t \cf}^{(0)}\, g_{\cf \cf}^{{(0)}}\, F_{t \cf} + g_{\cf \cf}^{(0)2}\, F_{\cf \cf}\Big], \\
    &g_{rr}^{(1)}= -\frac{\rho^2}{\Delta}\, f_{rr},\quad g_{\th \th}^{(1)}= - \rho^2\, h_{\th \th},\\
    &g_{\th \cf}^{(1)}= - g_{t \cf}^{(0)}\, h_{t \th} - g_{\cf \cf}^{(0)}\, h_{\th \cf}, \\
    &g_{t\th}^{(1)}= - g_{t t}^{(0)}\, h_{t \th} - g_{t \cf}^{(0)}\, h_{\th \cf}, \\
    &g_{tr}^{(1)}= -g_{tt}^{(0)}\, f_{tr} - g_{t \cf}^{(0)}\, f_{r \cf}, \\
    &g_{r \cf}^{(1)}= -g_{t \cf}^{(0)}\, f_{tr} - g_{\cf \cf}^{(0)}\, f_{r \cf}\, .
\end{split}
\end{equation}
Here, all $f$'s and $h$'s are functions of only $r$ and $\th$, respectively. Whereas $F_{ab}(r,\th) = f_{ab}(r)+h_{ab}(\th)$ for $\{a,b\} \in \{t,\cf\}$ and both $(f_{ab},h_{ab})$ are symmetric. Now, using the Appendix of Ref.~\cite{ Bezdekova:2022gib}, one can explicitly check that the above metric is in fact circular. This is also according to the result proven in Refs.~\cite{Delaporte:2022acp, Benenti:1979erw, Bezdekova:2022gib, Chen:2023gwm}, which dictate that geodesic separability implies circularity. At this point we must mention that though this result was known, we still went through the above exercise to make our analysis explicit and clear.\\

In order to make the metric non-circular, we must break geodesic separability. However, we shall do so in the simplest possible way, introducing a minimum (but necessary) number of beyond-Kerr modifications. But, before this, let us first impose two more conditions, namely asymptotic flatness and parameterized post-Newtonian (PPN) constraints, to make the above metric even simpler and easy to work with.

\subsection{Imposing Asymptotic Flatness and PPN constraints}
Since we are interested to study isolated BHs, asymptotic flatness is a very natural condition to use. Then, following the analysis of Ref.~\cite{Johannsen:2013szh}, we may express all non-zero $f$'s appearing in Eq.~\eqref{g1} as parameterized expansions:
\begin{equation}
\begin{split}\label{fh}
    &f_{\mu \nu}(r)= \sum_{n=0}^{\infty} \alpha_{\mu \nu}[n]\, \left(\frac{M}{r}\right)^n, 
\end{split}
\end{equation}
where the series coefficients $\alpha$'s are some constants depending on $M$, $a$, and possibly other hairs that might be present. Then, the asymptotic flatness requires at $r\to \infty$, the metric components to have the following forms: $g_{tt}=-1+2M/r+\mathcal{O}(r^{-2})$, $g_{t \cf}=-2 M a \sin^2\th/r+\mathcal{O}(r^{-2})$, $g_{rr}=1+2M/r+\mathcal{O}(r^{-2})$, $g_{\th \th} = \sin^{-2}\th\, g_{\cf \cf} = r^2+\mathcal{O}(r^{-1})$, and all other non-circular metric components to fall off as $1/r$ or faster. These structures are also useful since they help us identify the mass and spin of the BH.\\

For the metric in Eq.~\eqref{metric} with components given by Eq.~\eqref{g1}, the simplest 
choice of the series coefficients to assure asymptotic flatness is: $\alpha_{t \cf}[0-1]=\alpha_{\cf \cf}[0]=\alpha_{rr}[0-1]=\alpha_{tr}[0]=\alpha_{r \cf}[0-2]=0$; and $h_{t \cf}(\theta) = h_{\cf \cf}(\th) = h_{\th \th}(\th) = h_{t \th}(\th) = h_{\th \cf}(\th) = 0$. Here, we have used the shorthand $\alpha_{\mu \nu}[i-j]$ to represent all $\alpha_{\mu \nu}$ from index $i$ to $j$.\\

The deviation metric can be further constrained in the PPN framework \cite{Johannsen:2013szh}. Strictly speaking, one may avoid using the PPN constraints, since they are obtained from observations outside horizonless celestial objects. And, the exterior metric outside these objects may be different from a BH metric due to the loss of uniqueness in modified gravity. However, motivated by the analysis of Ref.~\cite{Johannsen:2013szh}, we may still use it to obtain a simpler metric as it implies $h_{tt}(\th) = f_{\cf \cf}(r) = \alpha_{tt}[0]=0$. Here, we have neglected other choices for simplicity and to avoid fine-tuning of the parameters.\\

Now that we have simplified the beyond-Kerr metric, let us come back to the issue of geodesic separability and circularity mentioned earlier. We have pointed out that, in order to make the metric non-circular, we must break geodesic separability. And, the simplest way to so is by making all $\alpha$'s some functions of $\th$ that preserve the $\mathbb{Z}_2$-symmetry about the equatorial plane. In this way, we make sure that our beyond-Kerr metric breaks Carter-symmetry due to the underlying non-circularity of the metric.\\

\textit{Therefore, with these constraints, the final non-circular metric becomes $g_{\mu \nu} = g_{\mu \nu}^{(0)}+\epsilon\, g_{\mu \nu}^{(1)}$, where $g_{\mu \nu}^{(0)}$ is the Kerr metric components given by Eq.~\eqref{kerr} and $g_{\mu \nu}^{(1)}$ is the deviation metric with the following non-vanishing components:}
\begin{equation}
\begin{split}\label{g1asymp}
    &g_{tt}^{(1)}= -\frac{1}{\rho^2} \Big[g_{tt}^{(0)2}\, \tilde{f}_{tt}+2\, g_{t \cf}^{(0)}\, g_{tt}^{{(0)}}\, \tilde{f}_{t \cf}\Big], \\
    &g_{t\cf}^{(1)}= -\frac{1}{\rho^2} \Big[g_{t \cf}^{(0)}\, g_{t t}^{(0)}\, \tilde{f}_{tt}+ \left( g_{t \cf}^{(0)2}+g_{t t}^{(0)}\, g_{\cf \cf}^{{(0)}}\right) \tilde{f}_{t \cf}\Big], \\
    &g_{\cf \cf}^{(1)}= -\frac{1}{\rho^2} \Big[g_{t \cf}^{(0)2}\, \tilde{f}_{tt}+2\, g_{t \cf}^{(0)}\, g_{\cf \cf}^{{(0)}}\, \tilde{f}_{t \cf}\Big], \\
    &g_{rr}^{(1)}= -\frac{\rho^2}{\Delta} \tilde{f}_{rr},\quad g_{\th \th}^{(1)}=g_{\th \cf}^{(1)}= g_{t\th}^{(1)}= 0, \\
    &g_{tr}^{(1)}= -g_{tt}^{(0)} \tilde{f}_{t r} - g_{t \cf}^{(0)} \tilde{f}_{r \cf}, \\
    &g_{r \cf}^{(1)}= -g_{t \cf}^{(0)} \tilde{f}_{tr} - g_{\cf \cf}^{(0)} \tilde{f}_{r \cf} ,
\end{split}
\end{equation}
\textit{where the non-vanishing functions $\tilde{f}_{\mu \nu}$ are given by}
\begin{equation}
\begin{split}\label{fhsymp}
    &\tilde{f}_{tt}(r,\th)= \sum_{n=1}^{\infty} \alpha_{tt}[n](\th)\, \left(\frac{M}{r}\right)^n ,\\
    &\tilde{f}_{t \cf}(r,\th)= \sum_{n=2}^{\infty} \alpha_{t \cf}[n](\th)\, \left(\frac{M}{r}\right)^n ,\\
    &\tilde{f}_{rr}(r)=f_{rr}(r)= \sum_{n=2}^{\infty} \alpha_{rr}[n]\, \left(\frac{M}{r}\right)^n ,\\
    &\tilde{f}_{t r}(r,\th)= \sum_{n=1}^{\infty} \alpha_{t r}[n](\th)\, \left(\frac{M}{r}\right)^n ,\\
    &\tilde{f}_{r \cf}(r,\th)= \sum_{n=3}^{\infty} \alpha_{r \cf}[n](\th)\, \left(\frac{M}{r}\right)^n .
\end{split}
\end{equation}
\textit{We also require that $\alpha_{rr}[n]$ to be $\theta$-independent (this is needed for the finiteness of the Kretschmann scalar at the horizon and the validity of the zeroth law of BH mechanics, discussed in the next section), and $\partial_\th \alpha_{\mu \nu}[n](\frac{\pi}{2}) =0$ to preserve the equatorial $\mathbb{Z}_2$-symmetry.}\\

We note that all $\alpha$'s except $\alpha_{rr}$'s and $\alpha_{tr}$'s are dimensionful. From the metric, one can easily read off the dimensions as: $[\alpha_{tt}] = [L^2]$, $[\alpha_{t \cf}] = [L]$, $[\alpha_{r \cf}] = [L^{-1}]$. Moreover, it is easy to check that the final metric $g_{\mu \nu}$ is not geodesic-separable and non-circular as the two expressions in Eq.~\eqref{circ} gives,
\begin{equation}
\begin{split}\label{circcheck}
&\xi \wedge \chi \wedge \dd \xi= \epsilon\, \frac{\Delta\, \sin^2\th}{\rho^2}\, \left[ (\cdots)\, \partial_\th \tilde{f}_{r \cf} + (\cdots)\, \partial_\th \tilde{f}_{tr} \right]\, \dd V, \\
&\xi \wedge \chi \wedge \dd \chi = \epsilon\, \frac{\Delta\, \sin^4\th}{2\rho^2}\,\left[ (\cdots)\, \partial_\th \tilde{f}_{t r} - (\cdots)\, \partial_\th \tilde{f}_{r \cf} \right]\, \dd V,
\end{split}
\end{equation}
where $\dd V = \dd t \wedge \dd r \wedge \dd \th \wedge \dd \cf$ and the ellipses contain some functions of $(r, \th)$ whose particular forms are not of particular use for us. Note that both of the above expressions are zero for Kerr ($\epsilon \to 0$), and for the special case $(\tilde{f}_{tr},\, \tilde{f}_{r \cf})$ are independent of $\th$. In contrast, for our case, both of the above expressions are non-zero and hence, the spacetime is non-circular.\\ 

However, by construction, our metric is $\mathbb{Z}_2$ symmetric. In fact, since $\partial_\th g_{\mu \nu}(\pi/2) = 0$, one can explicitly check that for all timelike/null geodesics with initial conditions $\{\th(\lambda =0) = \pi/2,\, \dot{\th}(\lambda=0)=0\}$, the acceleration $\ddot{\th}(\pi/2, \lambda \geq 0)$ vanishes trivially. Hence, the equatorial motion is stable similar to the Kerr case. \textit{Note, we shall keep denoting $\alpha_{\mu \nu}[n](\pi/2) := \alpha_{\mu \nu}[n]$ as a shorthand.} 

\section{Properties of the metric and geodesic equations}

Before we study the geodesic equations, let us first note some of the useful properties of the non-circular BH given by $g_{\mu \nu} = g_{\mu \nu}^{(0)}+\epsilon\, g_{\mu \nu}^{(1)}$, where the non-vanishing components of $g_{\mu \nu}^{(1)}$ are given by Eq.~\eqref{g1asymp}. For example, it is easy to check that various curvature scalars, such as Ricci scalar, Kretschmann scalar  etc, are regular in the domain of outer communication. Though we shall avoid writing their explicit expressions\footnote{Symbolically, $R = \epsilon\, \rho^{-5}\, (\cdots)$, $R_{\alpha \beta \gamma \delta}R^{\alpha \beta \gamma \delta} = \rho^{-6}\, (\cdots) + \epsilon\, \rho^{-10}\, (\cdots)$, and $R_{\alpha \beta}R^{\alpha \beta} = \Ord(\epsilon^2)$. The ellipses contain terms proportional to $\tilde{f}_{\mu \nu}$ and their $r$ and $\th$ derivatives up to 2nd order.}, as they are huge and not particularly illuminating! However, the most important among these aspects is to study the features of event horizon and ergosphere. 

\subsection{Event horizon and ergosphere}

The event horizon of any stationary and axisymmetric spacetime is defined to be the compact level-surface of the scalar function $H(r,\th)=r-h(\th)$ with a null normal $n_\mu = H_{,\mu}$. Then, at the event horizon, we must have $g^{rr}+g^{\th \th}\, h'(\th)^2 = 0$. Since our metric is $\mathbb{Z}_2$ symmetric, the second term vanishes at $\th = \pi/2$ and a good guess for the horizon location would be $g^{rr}(r_H) = 0$. In fact, using Eq.~\eqref{g1asymp}, one can explicitly check that $r_H = M+\sqrt{M^2-a^2}$ is indeed the event horizon, same as that of a Kerr BH. Moreover, this is also a Killing horizon where the Killing field $k^\mu = \xi^\mu + \Omega_H\, \chi^\mu$ becomes null. Here, $\Omega_H=a/(2\, M\, r_H)$ is the horizon's angular velocity and its value is same as that of a Kerr BH with the same mass $M$ and spin $a$. That is, the horizon of this BH rotates rigidly even though the metric is non-circular.\\

However, the constancy of $\Omega_H$ is not a mere coincidence. In fact, it is a direct consequence of the fact that for our metric, the both expressions in Eq.~\eqref{circcheck} are proportional to $\Delta(r)$ that vanishes at the horizon $r_H$. However, it does not imply that the underlying spacetime is circular, since for circularity $\xi \wedge \chi \wedge \dd \xi$ and $\xi \wedge \chi \wedge \dd \chi$ must vanish everywhere. One can also check that the location of the ergosphere is also same as that of Kerr, namely at $r_e=M+\sqrt{M^2-a^2\, \cos^2\th}$. An obvious way to check this is by noticing that $g_{tt}^{(1)}(r)$ vanishes at $r_e$.\\

From the above results, it may seem that all properties of Kerr horizon/ergosphere remains the same for our BH metric. However, let us emphasize that is not the case. For example, one can calculate the surface gravity $\kappa_H=\sqrt{-(k_{\mu;\nu})(k^{\mu;\nu})/2}$ at the event horizon \cite{Waldbook}, whose value turns out to be $\kappa_H = \kappa_H^{(0)}\, [1+\epsilon\, \tilde{f}_{rr}(r_H)/2]$. Here, $\kappa_H^{(0)} = \sqrt{M^2-a^2}/(2\, M\, r_H)$ is the surface gravity of a Kerr BH. Though the value of $\kappa_H$ may differ from that of Kerr (if $\tilde{f}_{rr}(r_H) \neq 0$), it is still a constant on the horizon. Hence, the zeroth law of BH thermodynamics holds true for these BHs \cite{Bardeen:1973gs, Waldbook} (this is the reason we earlier forced $\alpha_{rr}[n]$ to be $\th$-independent), making them good thermodynamic candidates in equilibrium. Moreover, all these novel properties make our BH metric a natural extension of the Kerr spacetime beyond circularity. In other works like in Refs.~\cite{Anson:2020trg, Anson:2021yli}, the reported non-circular metric suffers from many difficulties, such as the location of the event horizon is not a Killing horizon and its location varies with $\theta$.

\subsection{Geodesic equations}
Now, we are in a position to analyze the geodesic equations in the spacetime $g_{\mu \nu} = g_{\mu \nu}^{(0)}+\epsilon\, g_{\mu \nu}^{(1)}$, where the non-vanishing components of $g_{\mu \nu}^{(1)}$ are given by Eq.~\eqref{g1asymp}. Note that, due to the $\mathbb{Z}_2$-symmetry of the metric, the motion in the equatorial plane is stable and for our purpose, we shall confined ourselves only to equatorial geodesics parameterized by two constants of motion, namely the energy $E$, and $z$-component of the angular momentum $L_z$. We will also express the geodesic equations in such a way that visualize the deviations from that of Kerr in the most apparent way.\\

First of all, the equatorial $\dot{t}$ and $\dot{\cf}$ equations follow directly as a consequence of stationarity and axisymmetry of the spacetime. Additionally, using Eq.~\eqref{metricgeo}, one obtains the equatorial timelike/null geodesic equations up to the linear order in $\epsilon$ as follows:
\begin{equation}
\begin{split}\label{allgeo}
    &\dot{t}(\pi/2)=\dot{t}_{(0)}(\pi/2)-\frac{\epsilon}{r^2} \left(\tilde{E}\, \tilde{f}_{tt}-\tilde{L}_z\, \tilde{f}_{t \cf} \right)+\epsilon\, \tilde{f}_{tr}\, \dot{r}_{(0)}\, ,\\
    & \dot{\cf}(\pi/2)=\dot{\cf}_{(0)}(\pi/2)-\frac{\epsilon}{r^2}\, \tilde{E}\, \tilde{f}_{t \cf} + \epsilon\, \tilde{f}_{r \cf}\, \dot{r}_{(0)}\, , \, r^4\, \dot{\th}^2 = 0\, ,\\
    &r^4\, (1-\epsilon\, \tilde{f}_{rr})\, \dot{r}^2(\pi/2)=r^4\, \dot{r}_{(0)}^2-\epsilon\, \Delta\, \tilde{E} \left[\tilde{f}_{tt}\, \tilde{E}-2\, \tilde{f}_{t \cf}\, \tilde{L}_z\right],
\end{split}
\end{equation}
where $\dot{x}^\mu_{(0)}$ denotes the corresponding quantities for the Kerr case, the functions $\tilde{f}_{\mu \nu}$'s are given by Eq.~\eqref{fhsymp}, and $(\tilde{E},\, \tilde{L}_z)$ are the conserved energy and z-component of the angular momentum of a photon or quantities per unit mass for a massive particle. \textit{We must remember to evaluate the RHS of above expressions at the equatorial plane.} And, the timelike and null geodesics are differentiated by the fact that $\dot{x}^\mu_{(0)}$ (Kerr case) are different for these two cases.\\

From the $\dot{\th}$ equation, it is clear that the geodesic motion in the equatorial plane is stable, as discussed earlier. Moreover, we must note that the additional cross-components of the metric, such as $g_{tr}$ and $g_{r\cf}$, only enter into the $\dot{t}$ and $\dot{\cf}$ equations. Surprisingly, their effects cancel out in the $\dot{r}$ and $\dot{\th}$ equations. However, all equations (except that of $\dot{\th}$) are still modified from the corresponding Kerr case due to the presence of non-circularity parameter $\epsilon$.

\subsection{Circular Orbits}

Among all geodesics, circular ones are of particular interests for their observational relevance. For example, circular photon orbits are closely linked with the BH shadow formation, whereas the circular timelike ones are important for binary dynamics and accretion studies. For our case, since the equatorial motion is stable, circular orbits are confined to this plane and we need to only consider the $\dot{r}$ equation in Eq.~\eqref{allgeo}. In particular, this equation can be rewritten as $g_{rr}\dot{r}^2+V(r)=0$, where the effective potential $V(r)$ has to satisfy $V(r)=V'(r)=0$ for circular orbits. Putting in all the necessary ingredients, one can rewrite the potential as
\begin{equation}
\begin{split}\label{circle}
    V(r) = \delta+g^{tt}\, \tilde{E}^2-2\, g^{t \cf}\, \tilde{E}\, \tilde{L}_z+g^{\cf \cf}\, \tilde{L}_z^2+\mathcal{O}(\epsilon^2)\ ,
\end{split}
\end{equation}
where, from Eq.~\eqref{g1asymp}, one can easily find out that at the equatorial plane: $g^{tt} = g^{tt}_{(0)}(\th=\pi/2)+\epsilon\, \tilde{f}_{tt}(\th=\pi/2)/r^2$, $g^{t\cf} = g^{t\cf}_{(0)}(\th=\pi/2)+\epsilon\, \tilde{f}_{t\cf}(\th=\pi/2)/r^2$, and $g^{\cf \cf} = g^{\cf \cf}_{(0)}(\th=\pi/2)$. Also, the parameter $\delta = 1$ ($0$) if the geodesic is timelike (null).\\

For equatorial circular photon orbits (light rings), the form of the potential looks similar to that given in Ref.~\cite{Cunha:2017qtt}. Thus, although their results regarding the existence of light rings were derived assuming the spacetime to be circular \cite{Cunha:2020azh, Guo:2020qwk}, they are still valid for our non-circular metric too. Intuitively, this is because the extra cross-terms in the metric do not affect the Lagrangian $2 \mathcal{L}=g^{\mu \nu}\, p_\mu p_\nu$ for a circular orbit with $p_r=p_\th=\dot{p}_r=\dot{p}_\th=0$ (condition for light rings). Hence, outside the event horizon, our rotating BH must have an odd number of light rings for each rotation sense. In fact, following Ref.~\cite{Ghosh:2021txu}, we can further show that at least one light ring exists outside the ergoregion. Note that due to the reflection symmetry of the metric about the equatorial plane, light rings will exist at $\theta = \pi/2$. Then, the radial locations ($r_\gamma^{\pm}$) of the prograde (upper sign)/retrograde (lower sign) light rings can be found by solving $p_r=\dot{p}_r=0$. However, since for our purpose, finding the explicit locations is not necessary, we shall only quote the result up to the quadratic order in $(a/M)$,  
\begin{equation}
\begin{split}\label{lrs}
r_\gamma^{\pm} \approx &\, 3 M \mp \frac{2 a \sqrt{3}}{3} - \frac{2 a^2}{9 M} \mp \epsilon \frac{a \tilde{f}_{t t}}{243 M^3} (3\sqrt{3} M \pm 2 a) \\
& \pm \epsilon \frac{2 a \tilde{f}_{t \cf}^{\pm} }{9 M} + \epsilon\frac{\partial_r \tilde{f}_{t t}}{162 M^2}(-9 M^2 \pm 4 a \sqrt{3} M + a^2) \\
& + \epsilon \frac{\sqrt{3}\, \partial_r \tilde{f}_{t \cf}^{\pm}}{162 M}(54 M^2 \mp 36 a \sqrt{3} M + 7 a^2)\, ,
\end{split}
\end{equation}
where we have introduced the notations $\tilde{f}_{t \cf}^\pm = \pm \tilde{f}_{t \cf}$, which are evaluated at the equatorial plane and at the location of the prograde/retrograde Kerr light ring ($\epsilon = 0$) keeping terms up to $(a/M)^2$ together with the factors multiplying them. We should also note that among these light rings, the retrograde one ($r_\gamma^{-}$) remains outside the ergoregion all allowed values of spin $0<a<M$; whereas the prograde one ($r_\gamma^{+}$) enters the ergoregion for near-extremal spin values.\\

Now, let us move on to the discussion of the equatorial timelike circular orbits, which will be useful in the subsequent sections. These orbits are characterized by three orbital parameters $\{\tilde{E}, \tilde{L}_z, \Omega\}$ with $\Omega=\dot{\cf}/\dot{t}$ being the angular velocity of the orbiting test particle. Then, from Eq.~\eqref{circle} with $\delta = 1$, we can solve $V(r)=V'(r)=0$ to get (all quantities in the RHS are evaluated at $\th = \pi/2$),
\begin{equation}
\begin{split}\label{elqm}
    &\tilde{E}=\frac{-g_{t t}-g_{t \cf}\, \Omega}{\sqrt{-g_{t t}-2 g_{t \cf}\, \Omega-g_{\cf \cf}\, \Omega^2}}\, , \\
    &\tilde{L}_z= \frac{g_{t \cf}+g_{\cf \cf}\, \Omega}{\sqrt{-g_{t t}-2 g_{t \cf}\, \Omega-g_{\cf \cf}\, \Omega^2}}\, , \\
    &\Omega_{\pm}=\frac{-\partial_r g_{t \cf} \pm \sqrt{\left(\partial_r g_{t \cf}\right)^2-\partial_r g_{t t}\, \partial_r g_{\cf \cf}}}{\partial_r g_{\cf \cf}}\, .
\end{split}
\end{equation}
Here, the upper (lower) sign corresponds to the prograde (retrograde) motions. The explicit expressions of the conserved quantities $(\tilde{E}, \tilde{L}_z)$ are given in Appendix-\ref{app:cons}. However, from the above derivation it may not be clear that $\Omega$ is indeed $\dot{\cf}/\dot{t}$, where $\dot{\cf}$ and $\dot{t}$ are calculated for a circular orbit using Eq.~\eqref{allgeo}. To see why this holds, it suffices to note that for a circular orbit, the Euler-Lagrange equation is simply $\partial_r \mathcal{L} = 0$, since $d[\partial_{\dot{r}} \mathcal{L}]/d\lambda = 0$ (even for non-circular spacetime). This, in turn, implies $\partial_r g_{tt}+2\, \Omega\, \partial_r g_{t \cf}+\Omega^2\, \partial_r g_{\cf \cf}=0$, whose roots are clearly $\Omega_\pm$ given above.\\

For studies regarding both binary dynamics and accretion physics, one particularly important timelike orbit is the so-called innermost stable circular orbit (ISCO). It is a marginally stable orbit that marks the end of the allowed region for timelike circular orbits. To find its location $r=r_{ISCO}$, we need to solve for the smallest positive root of the equation $V''(r,\tilde{E},\tilde{L}_z)=0$, where the effective potential is given in Eq.~\eqref{circle} and $(\tilde{E},\tilde{L}_z)$ are given by Eq.~\eqref{elqm}. Now, if one writes as $V''=V_{(0)}''+\epsilon\, V_{(1)}''$ and $r_{ISCO}=r_{(0)}+\epsilon\, r_{(1)}$, then one needs to solve for the smallest real $r_{(1)}$ solving $V_{(0)}'''(r_{(1)})=-V_{(1)}''(r_{(0)})$. Here, we have assumed that $r_{(0)}$ is the Kerr ISCO radius. However, since we do not need the exact expression of the ISCO radius in this work, we shall skip deriving it and move on to discussing various observable effects of non-circularity.

\section{Observable effects of non-circularity} 

In the previous section, we have studied several mathematical properties of our non-circular spacetime. Apart from these theoretical studies, we shall now discuss possible ways to constrain the non-circular deviations from observations. For this purpose, we shall focus on the studies of EMRIs and LT effect, which proves to be very useful for constraining various model-parameters. For example, as discussed in the introduction section, EMRI framework will help us capture the effect of non-circular deviations from Kerr paradigm by providing a thorough map of the spacetime outside the primary SMBH -- a promising prospect for the future GW observations by LISA.\\ 

Apart from EMRI, we shall see that LT effect will provide a powerful way to constrain the $\epsilon\, \alpha_{t \cf}[2]$ term in the metric, thereby restricting the strength of the non-circularity paramter $\epsilon$. Let us also point out that by construction the appearance of $f_{tt},\,  f_{t\cf}$, and other beyond-Kerr terms are supported by the underlying non-circularity of the metric. Moreover, since the geodesically-conserved energy and angular momentum depend explicitly on $(f_{tr},\, f_{r\cf})$, all measurements/observations will be inevitably influenced by non-circularity. 

\subsection{Study of EMRI} \label{ObsEMRI}
For the study of EMRI in our non-circular SMBH spacetime, we shall implement the ``hybrid waveform'' scheme extensively used in the literature, see for example Refs.~ \cite{Glampedakis:2005cf, Babichev:2024hjf, Sopuerta:2009iy, Pani:2011xj, Canizares:2012is, Chua:2017ujo, Destounis:2021mqv, Collodel:2021jwi, Delgado:2023wnj} and references therein. It is a very powerful tool to gauge the non-Kerr effects in GW emission from a binary. In this method, one takes a minimalist's approach by assuming the binary looses energy/angular momentum through GW emission modeled by the well-known Einstein quadrupole formula and thereby neglects other dissipative effects arising from spin, higher curvature or matter fields. The dissipation causes the circular orbit of the secondary shrinks adiabatically. This framework adequately captures the leading order post-Newtonian dissipative effects \cite{Glampedakis:2005cf, Babichev:2024hjf}. It should be mentioned here that our method inherits the drawback of the hybrid waveform scheme. We cannot predict the evolution of the binary in an absolute sense, because the post-Newtonian modelling of GW emission is  perturbative in spin resulting in the quadrupole formula to be independent of spin effects at leading order. Nevertheless, the hybrid waveform method still allows one to compare and contrast the non-circular spacetime from Kerr, and understand the nature of deviations qualitatively and quantitatively at an order-of-magnitude level.\\

\begin{table}[h!]
\centering
{
\renewcommand{\arraystretch}{1.4}
\begin{tabular}{||c|c|c|c|c|c||} 
 \hline
 Model & $\epsilon$ & $M^{-2} \alpha_{tt}[1]$ & $M^{-2} \alpha_{tt}[2]$ & $M^{-1} \alpha_{t\phi}[2]$ & $M^{-1} \alpha_{t\phi}[3]$ \\
 \hline\hline
Kerr & 0 & 0 & 0 & 0 & 0 \\
\hline
I & 0.08 & $\frac{1}{5}$ & $\frac{1}{10}$ & $\frac{1}{10\sqrt{10}}$ & $\frac{1}{20\sqrt{10}}$ \\
 \hline
 II & 0.08 & $-\frac{1}{5}$ & $\frac{1}{10}$ & $-\frac{1}{10\sqrt{10}}$ & $\frac{1}{20\sqrt{10}}$ \\
 \hline\hline
\end{tabular}
}
\caption{Parameter choices for Model I and II. In both cases, the non-circular deviation parameter is chosen to be $\epsilon = 0.08$.}
\label{tab:model}
\end{table}

Following this framework, we have considered two distinct beyond-Kerr models and labelled them as Model I and Model II. These models correspond to different choices of the non-circular deviation parameters $\tilde{f}_{tt}(r,\pi/2)$ and $\tilde{f}_{t\phi}(r,\pi/2)$, as defined in Eq.~\eqref{fhsymp}. Before highlighting the explicit choices, it is important to understand the motivation behind those specific choices. Owing to our perturbative expansions around Kerr, the quantities $g_{tt}^{(1)}$ and $g_{t\phi}^{(1)}$ must be small corresponding to their background values. From Eq.~\eqref{g1asymp}, this implies that $\epsilon\, |\alpha_{tt}|/M^2 << 1$ and $\epsilon\, |\alpha_{t\phi}|/M << 1$ (the factors of $M$ are to make the associated quantities dimensionless). Additionally, we see that terms containing higher $\alpha_{\mu \nu}[n]$ are suppressed by the factor of $r^{-n}$. Therefore, for our purpose we have assumed the only non-zero $\alpha$'s to be $\{\alpha_{tt}[1],\, \alpha_{tt}[2]\}$ and $\{\alpha_{t\phi}[2],\, \alpha_{t\phi}[3]\}$. The explicit choices of parameters of these models are highlighted in Tab.-\ref{tab:model}.\\ 

\begin{figure}[ht!]
    \centering
    \includegraphics[width=1.0\columnwidth]{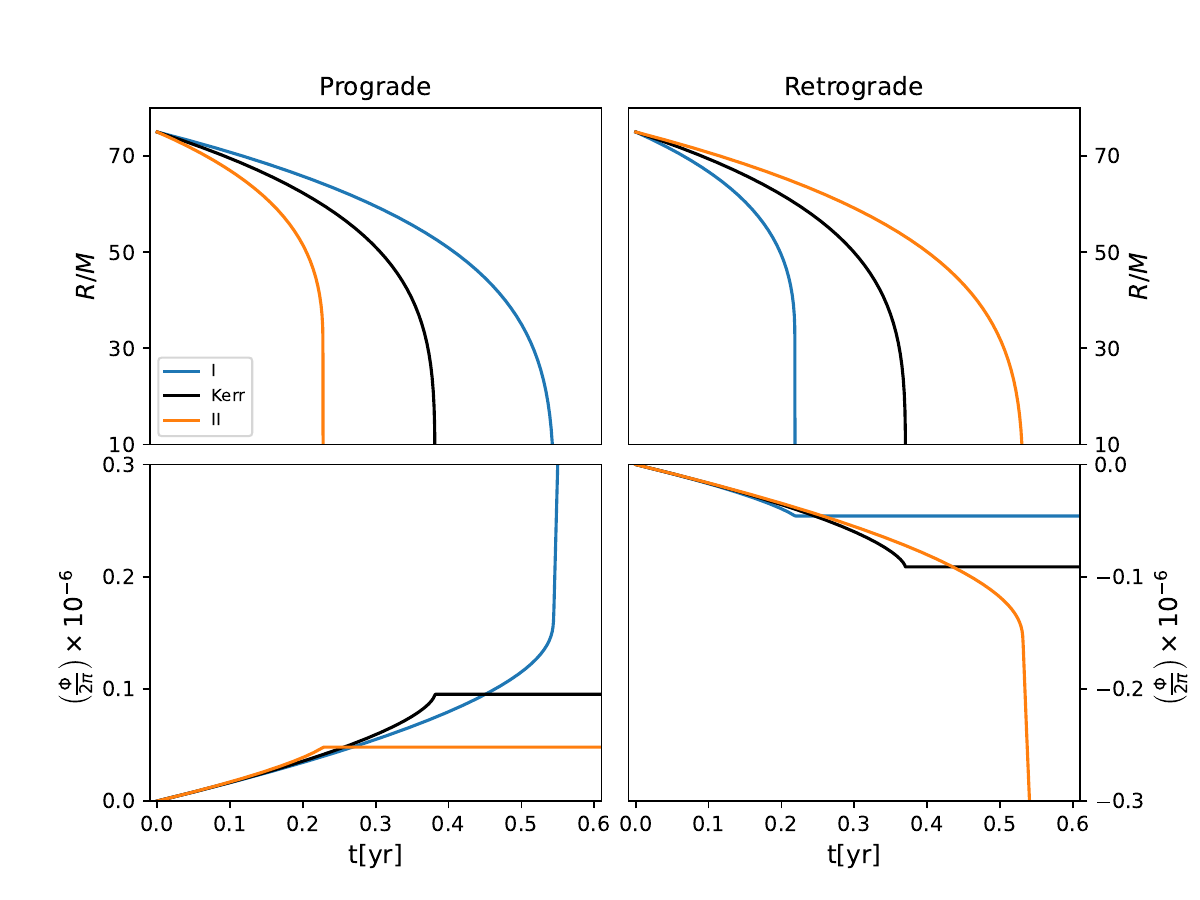}
    \caption{Plots showing the adiabatic shrinkage of the orbital radius $R$ (upper panels) and the cumulative number of orbits (lower panels), for the two Models I and II along with the Kerr case. The time-ticks are shown as a fraction of a year. The flattening feature seen on the lower panels is because of the cumulative phase becoming constant as the simulation is terminated below $10M$ to maintain numerical accuracy.}
    \label{fig:ROmega}
\end{figure}

Let us emphasize that our choice of parameters from Tab.-\ref{tab:model} makes $g_{\mu\nu}^{(1)}/g_{\mu\nu}^{(0)} \lesssim 0.016$ at the equatorial plane, which is truly a tiny fraction. Interestingly, it is seen that even such a small non-circular deviation produces noticeable effects, as demonstrated in Fig.~\ref{fig:ROmega}. We have chosen the mass and spin of the primary SMBH to be $M = 10^4 M_\odot$ and $a/M = 0.5$, respectively. Additionally, the secondary object is assumed to have a mass of $m=25 M_\odot$. Since our analysis treats the secondary as a test object orbiting around the central SMBH, we do not require to fix the nature of secondary, $\ie$, it could be a BH or horizonless object.\\

Then, from Fig.~\ref{fig:ROmega} (see in the upper panels) that plots the adiabatic shrinkage of the orbital radius\footnote{This choice of circumferential radius is both geometrical and natural, and motivated from the analysis of Ref.~\cite{Babichev:2024hjf}.} $R:=\sqrt{g_{\cf \cf}(\th=\pi/2)}$, we observe that the merger timescale for our beyond-Kerr models is almost double (half) w.r.t Kerr for prograde (retrograde) motion. This outcome has crucial implications on the cumulative orbital phasing, which is evident from the lower panels of Fig.~\ref{fig:ROmega}.\\

\begin{figure}[h!]
    \centering
    \includegraphics[width=1.0\columnwidth]{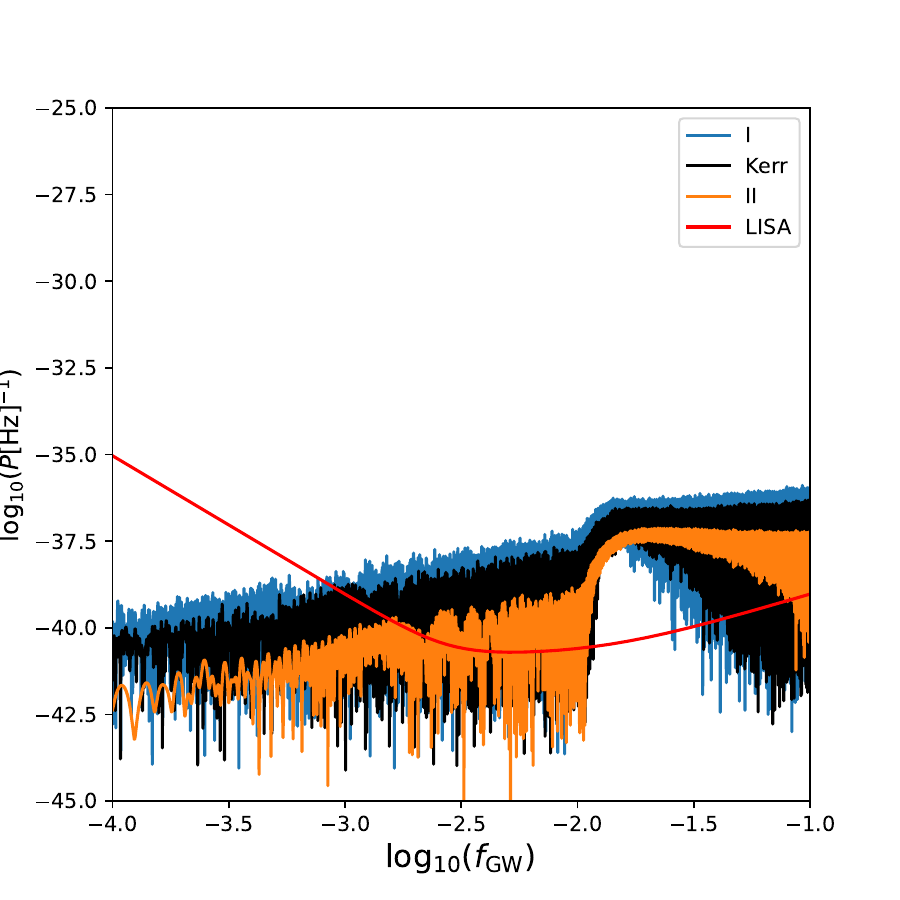}
    \caption{The LISA noise power plotted (in red) as a function of the GW frequency $f_{\text{GW}}$. The signal power for Kerr along with models I and II are also shown. All signals start around $0.01$ Hz, and would be clearly visible in the LISA band.}
    \label{fig:PSD}
\end{figure}

Despite the striking deviations from the Kerr paradigm, we must point out that the merger timescale is not a direct GW observable. Hence, we move on to evaluating a more relevant quantity for observations, namely the \textit{signal to noise ratio} (SNR). For this purpose, we recall the asymptotic form of the GW strain for a face-on binary as \cite{Maggiore:2007ulw},

\begin{align}\label{eq:hphc}
    h_+(t) &= 4\left(\frac{m}{D_l}\right) \left(\frac{M}{R(t)}\right) \cos[2\Phi(t)] ,\nonumber \\
    h_\times(t) &= 4\left(\frac{m}{D_l}\right) \left(\frac{M}{R(t)}\right) \sin[2\Phi(t)] ,\nonumber \\
\end{align}
where $D_l$ is the luminosity distance and $\Phi(t)$ is the cumulative orbital phasing plotted in the lower panels of Fig.~\ref{fig:ROmega}. We have also utilized the fact that the phase of quadrupolar GW radiation is twice of the orbital phase. Now, using Eq.~\eqref{eq:hphc}, one can calculate the signal spectral power in the frequency domain, and then compare it with the noise spectral power of LISA \cite{Sathyaprakash:2009xs}.\\ 

The above Fig.~\ref{fig:PSD} shows the power spectral densities of the signals for Model I, Model II and the Kerr case along with the LISA detector noise has for the prograde orbits (a similar plot can be generated for retrograde motion as well).  Apart from the choices of $\{M,m,\chi\}$ from before, we also consider a luminosity distance $D_l = 400$ Mpc. From Fig.~\ref{fig:PSD}, it is clear that all signals irrespective of the models would be clearly visible and distinguishable in the LISA band.\\

\begin{table}[h!]
\begin{center}
\begin{tabular}{||m{1.5cm}| m{1.5cm}| m{1.5cm}| m{1.5cm}||} 
 \hline
 Model & $f_l$[Hz] & $f_u$[Hz] & SNR \\ [0.5ex] 
 \hline\hline
 Kerr & $10^{-5}$ & $0.1$ & 68.2 \\ 
 \hline
 I & $10^{-5}$ & $0.1$ & 89.8 \\ 
 \hline
 II & $10^{-5}$ & $0.1$ & 41.3 \\ [1ex] 
 \hline
 \hline
\end{tabular}
\end{center}
\caption{We mention the lower and upper cut off frequencies $f_l$ and $f_u$, respectively. The associated SNR's for different models are also quoted.}
\label{tab:lusnr}
\end{table}

As a final measure of distinction, we have listed the numerical values of the SNR's in Tab.-\ref{tab:lusnr}. It is immediately clear that the in-band SNR's change by almost a factor of two between model I and II and a factor of $1.3-1.6$ from the Kerr case. The striking difference in SNR's  highlights that the future parameter estimation studies performed on these models have an excellent chance of inferring the beyond-Kerr parameters, namely $\epsilon$ and $\alpha$'s.
  
\subsection{Study of LT precession}
The EMRI framework presented in the previous subsection can only constraint some combinations of $\epsilon$ and $\alpha$ parameters. However, we shall now see that the future detection of strong-field LT precession can be used to specifically constraint the $\alpha_{t \cf}[2]$ term.\\

The LT precession frequency is defined relative to an orthonormal tetrad frame $e_\alpha$ that is Lie-dragged along the integral curves of the timelike (outside the ergoregion) Killing vector field $\xi$. Such a frame is known as the \textit{Copernican frame}, in which $e_\alpha$ can be interpreted as the axes at rest about which a gyroscope precesses due to frame dragging. Then, following the analysis of Refs.~\cite{Straumann:2013spu, Lightman2017:2017mmh}, the exact form of the LT $4$-frequency is given by $\Omega_\mu = \frac{1}{2\xi^2} \eta_\mu^{\, \, \, \nu \rho \sigma}\xi_\nu \partial_\rho \xi_\sigma$, where $\eta$ represents the volume-form in the spacetime. Hence, a test gyroscope will not precess if and only if $\xi \wedge d \xi = \eta_\mu^{\, \, \, \nu \rho \sigma}\xi_\nu \partial_\rho \xi_\sigma$ vanishes identically.\\

In a Kerr spacetime, this frame dragging is essentially due to presence of the cross-term $g_{t \cf}$ in the metric that signifies rotation of the central object. However, in a general stationary spacetime expressed in terms of coordinates adapted to the timelike Killing isometry, the co-vector $\xi_\alpha = g_{\alpha 0}$. Thus, in the chosen frame of reference, the spatial components of $\Omega_\mu$ can be written in the co-vector form as~\cite{Straumann:2013spu}, 
 \begin{equation}\label{LTco}
    \Omega_l =  \frac{\epsilon_{ijl}}{2\sqrt{-g}} \left[ g_{0i,j} \left( \partial_l -\frac{g_{0l}}{g_{00}} \partial_0 \right) - \frac{g_{0i}}{g_{00}}\, g_{00,j}\, \partial_l \right] \, , 
\end{equation}
where $\epsilon_{ijl}$ (Latin indices run over the spatial indices 1,2, and 3) is the fully-antisymmetric symbol and the repeated indices are summed over. However, we are only interested in the far-field limit of the magnitude $\Omega_{LT}(r,\pi/2)$ of this vector. For our metric given by Eq.~\eqref{g1asymp}, it is now easy to calculate the magnitude of the equatorial LT frequency in the far-field form up to $\mathcal{O}(r^{-4})$ as,
\begin{equation}\label{LTo}
    \Omega_{LT}(r,\pi/2) \approx  \Omega_{LT}^{(0)}(r,\pi/2) - \epsilon\, \alpha_{t \cf}[2]\, \frac{M^2}{r^4} \, ,
\end{equation}
where its Kerr value $\Omega_{LT}^{(0)}$ is given by Eq.~\eqref{LTk} at $\th = \pi/2$. Note that the leading order cubic fall-off remains the same as the Kerr case, whereas the effect of non-circularity modifies the $r^{-4}$ and higher order terms. Therefore, in a weaker gravitational field like that of earth, the deviation due to non-circularity might remain hidden and becomes detectable only in the presence of strong gravity. In fact, as suggested in Ref.~\cite{Chakraborty:2013naa, Chakraborty: Texas}, such strong-gravity modifications in LT precession are a potential smoking gun for probing yet-unknown gravitational physics in the context of BH accretion and X-ray emission from neutron stars.\\

\begin{figure}[ht!]
    \centering
    \includegraphics[width=\columnwidth]{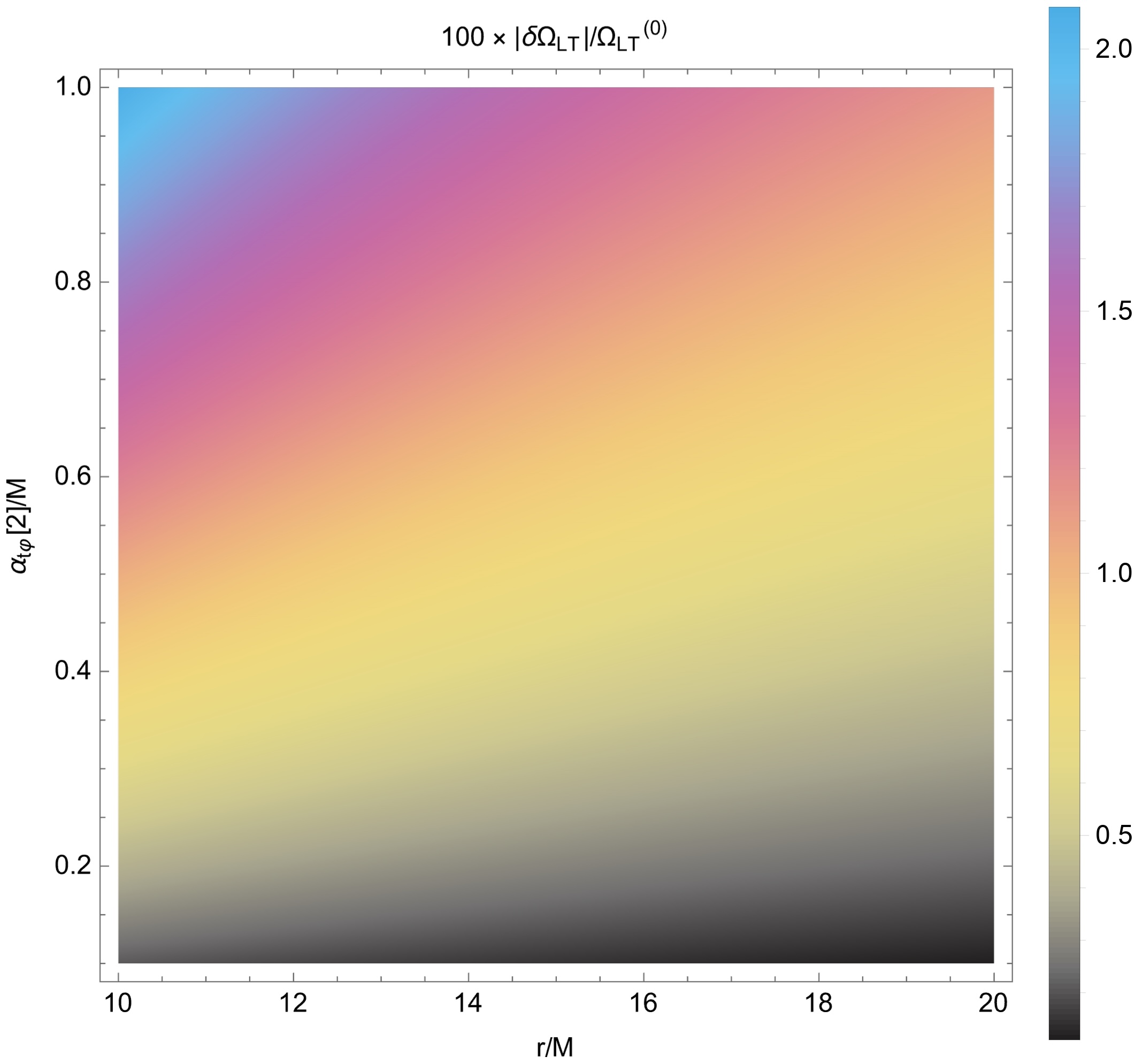}
    \caption{Beyond-Kerr variation to the LT precession frequency due to non-circularity for a fixed $a/M = 0.4$, $\th = \pi/2$, and $\epsilon=0.1$. The value of spin is based on that of the central object in Sgr A* \cite{EHT5, Kato:2010}. The side color-bar shows the relative percentage variation in $\Omega_{LT}$ from its equatorial Kerr value.}
    \label{fig:LTdev}
\end{figure}

Moreover, from the above equation, another important point to note is that the deviation term only depends on a single beyond-Kerr parameter $\alpha_{t \cf}[2]$. It gives us a direct way to constrain this parameter from future observations of strong-field LT precession. For a better visualization, let us plot the percentage change in the absolute value of $\Omega_{LT}$ from its equatorial ($\th = \pi/2$) Kerr value for the spin $a/M = 0.4$ and non-circularity parameter $\epsilon = 0.1$.\\

The above Fig.~[\ref{fig:LTdev}] clearly demonstrates that as one moves closer to the central object, which is taken to be that in Sgr A* for the purpose of the plot, the deviation due to non-circularity becomes more evident for a fixed $\alpha_{t \cf}[2]$. This feature offers an excellent opportunity to detect even a small non-circular deviation from the Kerr paradigm in future observations of strong-field LT effect. Moreover, with a better accuracy, such future measurements can be used to tightly constrain the $\alpha_{t \cf}[2]$ parameter.

\section{Summary and Discussions}
One of the primary objectives of both the GW observations by LVK collaboration \cite{LIGO1, LIGO2, LIGO3, LIGO4, LIGO5, LIGO6, LIGO7} and BH shadow imaging by the Event Horizon Telescope \cite{EHT1, EHT2, EHT3, EHT4, EHT5, EHT6} is to probe and constrain potential deviations from the Kerr paradigm. Detecting such deviations could provide crucial evidence of new gravitational physics beyond GR. Strong gravitational environments offer the most promising opportunities for this, as they can amplify such deviations like a magnifying glass. And, a possible method to detect post-Kerr effects is to examine the observational consequences (or lack thereof) of various Kerr symmetries, like circularity.\\

To study the upshots of non-circularity in a theory-agnostic and
unified manner, we have constructed a parameterized non-circular metric that is slightly deviated from that of Kerr. This new metric retains the other Kerr properties and clearly isolate the effects of non-circularity. Next, we move on to a rigorous discussion of various theoretical properties of this BH metric, including the location of the event horizon, ergosphere, the presence of light ring(s), and ISCO. All these quantities are expressed in such a way that explicitly highlights beyond-Kerr deviation. Moreover, the previous analysis has demonstrated that our non-circular BHs have a regular event horizon (which is also a killing horizon) with a constant angular velocity and they obey the zeroth law, making them natural candidates for studying the other three laws of BH mechanics beyond circularity. These novel properties are distinct from various existing non-circular metric in the literature~\cite{Anson:2020trg, Anson:2021yli}, warranting further investigations of our metric. A follow up on such important aspects is left as a future exercise.\\

To complete our analysis, we then explore various observational effects of non-circularity. In particular, we focus on the studies of equatorial EMRIs and LT effect, which have proven to be very useful for constraining
model parameters. Both of these studies have clearly demonstrated the promising prospects of detecting and constraining non-circular departures from the Kerr paradigm in future observations. In particular, the strong-field LT effect will offer a direct way to constrain the $\alpha_{t \cf}[2]$ parameter. Whereas if the spacetime outside a SMBH exhibits non-circular deviations, these can be effectively captured by LISA through the observations of EMRI signals. Specifically, in Sec.~\ref{ObsEMRI}, we have shown that all signals irrespective of
the underlying non-circular models would be clearly visible and distinguishable
in the LISA band from the Kerr paradigm.\\

In future, it will be interesting to include the effect of eccentricity in secular evolution of various orbital parameters. It will help us modeling the EMRI signals more accurately~\cite{Amaro-Seoane:2012lgq, Datta:2023uln} and also to avoid systematic biases in parameter estimation analyses~\cite{Favata:2021vhw}. Other possible future extensions of our work could include the study of perturbations and the associated quasi-normal modes of these non-circular BHs, examining tidal effects in the presence of external perturbations~\cite{Datta:2019euh, Datta:2019epe, Bernaldez:2023xoh, Datta:2024vll}, and investigating shadow signatures. These studies will further help constrain the non-circular deviations from the Kerr paradigm. 

\section*{Acknowledgements}
We are thankful to Che-Yu Chen and Hsu-Wen Chiang for many helpful discussions and pointing out to us the important connection between geodesic separability and circularity. The authors also want to thank Sudipta Sarkar for useful discussions and thoughtful comments on an earlier draft. The research of K.C is supported by the PPLZ grant (Project number: 10005320/ 0501) of the Czech Academy of Sciences.

\appendix

\section{The inverse metric}\label{app:inverse}
In this appendix, we explicitly write down the covariant metric components used in the main text. For this purpose, it is useful to first note the inverse Kerr metric components:
\begin{equation}
\begin{split}\label{kerrinv}
    &g^{tt}_{(0)}= - \frac{g_{\cf \cf}^{(0)}}{J},\, g^{t \cf}_{(0)}= \frac{g_{t \cf}^{(0)}}{J},\, g^{\cf \cf}_{(0)}= - \frac{g_{t t}^{(0)}}{J},\\
    &g^{r r}_{(0)}= \frac{1}{g_{r r}^{(0)}},\, g^{\th \th}_{(0)}= \frac{1}{g_{\th \th}^{(0)}}\, ,
\end{split}
\end{equation}
where $g_{\mu \nu}^{(0)}$ can be read from the Kerr metric given in Eq.~\eqref{kerr}, and $J = g_{t \cf{(0)}}^2-g_{t t{(0)}}\, g_{\cf \cf{(0)}}=\Delta\, \sin^2\th$. With these notations, the inverse components (up to linear order in $\epsilon$) of the metric in Eq.~\eqref{metric} are given by,
\begin{equation}
\begin{split}\label{metricinv}
    &g^{tt}= g^{tt}_{(0)} +\frac{\epsilon}{J^2} \Big[-g_{\cf \cf}^{(1)}\, g_{t \cf}^{(0)2}+2\, g_{t \cf}^{(0)}\, g_{\cf \cf}^{{(0)}}\, g_{t \cf}^{(1)} - g_{\cf \cf}^{(0)2}\, g_{t t}^{(1)}\Big], \\
    &g^{t\cf}= g^{t\cf}_{(0)} +\frac{\epsilon}{J^2} \Big[-g_{t \cf}^{(1)}\, g_{t \cf}^{(0)2}+ g_{t \cf}^{(0)}\, g_{t t}^{{(0)}}\, g_{\cf \cf}^{(1)} - g_{t t}^{(0)}\, g_{\cf \cf}^{{(0)}}\, g_{t \cf}^{(1)}\\
    &\hskip 18em +g_{\cf \cf}^{(0)}\, g_{t \cf}^{(0)}\, g_{t t}^{(1)}\Big], \\
    &g^{\cf \cf}= g^{\cf \cf}_{(0)} +\frac{\epsilon}{J^2} \Big[-g_{\cf \cf}^{(1)}\, g_{t t}^{(0)2}+2\, g_{t \cf}^{(0)}\, g_{tt}^{{(0)}}\, g_{t \cf}^{(1)} - g_{t \cf}^{(0)2}\, g_{t t}^{(1)}\Big], \\
    &g^{rr}= g^{rr}_{(0)}-\epsilon\, \left[\frac{g_{rr}^{(1)}}{g_{rr}^{(0)2}}\right],\quad g^{\th \th}= g^{\th \th}_{(0)}-\epsilon\, \left[\frac{g_{\th \th}^{(1)}}{g_{\th \th}^{(0)2}}\right],\\
    &g^{r \th}= -\frac{\epsilon\, g_{r\th}^{(1)}}{g_{rr}^{(0)}\, g_{\th \th}^{(0)}}\, ,\quad g^{\th \cf}= \frac{\epsilon}{J\, g_{\th \th}^{(0)}} \Big[g_{\th \cf}^{(1)}\, g_{t t}^{(0)} - g_{t \cf}^{(0)}\, g_{t \th}^{(1)}\Big], \\
    &g^{t\th}= \frac{\epsilon}{J\, g_{\th \th}^{(0)}} \Big[g_{t \th}^{(1)}\, g_{\cf \cf}^{(0)} - g_{t \cf}^{(0)}\, g_{\th \cf}^{(1)}\Big], \\
    &g^{tr}= \frac{\epsilon}{J\, g_{rr}^{(0)}} \Big[g_{t r}^{(1)}\, g_{\cf \cf}^{(0)} - g_{t \cf}^{(0)}\, g_{r \cf}^{(1)}\Big], \\
    &g^{r \cf}= \frac{\epsilon}{J\, g_{rr}^{(0)}} \Big[g_{r \cf}^{(1)}\, g_{tt}^{(0)} - g_{t \cf}^{(0)}\, g_{t r}^{(1)}\Big]\, .
\end{split}
\end{equation}
Note that, at this stage, all the metric components appearing in the above equations are functions of $(r,\theta)$. 

\section{Conserved quantities for timelike circular orbit}\label{app:cons}
In this appendix, we have listed down the conserved quantities, namely $(\tilde{E},\tilde{L}_z)$, for an equatorial timelike circular orbit. In particular, these quantities will be expressed in such a way that makes the non-circular deviation from Kerr apparent. For this purpose, let us first mention $(\tilde{E}^{(0)},\tilde{L}_z^{(0)})$ for the Kerr spacetime,
\begin{equation}
\begin{split}\label{consKerr}
&\tilde{E}^{(0)} = \frac{r^{3/2} - 2 M r^{1/2} \pm a M^{1/2}}{r^{3/4}\, \sqrt{Z_\pm}}, \\
&\tilde{L}_z^{(0)} = \pm \frac{M^{1/2} (r^2 \mp 2 a M^{1/2} r^{1/2} + a^2)}{r^{3/4}\, \sqrt{Z_\pm}}\, ,
\end{split}
\end{equation}
where we have $Z_\pm=\pm 2 a M^{1/2} - 3 M r^{1/2} + r^{3/2}$ and the upper (lower) sign is for prograde (retrograde) motion. Now, using Eq.~\eqref{elqm}, we can write down the conserved quantities associated with an equatorial timelike circular orbit in our non-circular spacetime: 
\begin{equation}
\begin{split}\label{consOur}
&\tilde{E}=\tilde{E}^{(0)} + \frac{\epsilon\, \tilde{E}^{(0)} \left[ \tilde{E}^{(0)}\, U_{tt}^\pm \pm 2\,\tilde{L}_z^{(0)}\, U_{t \cf}^\pm \right]}{4\, r^{13/4}\, \sqrt{Z_\pm}}, \\
&\tilde{L}_z=\tilde{L}_z^{(0)} \pm \frac{\epsilon\, \tilde{E}^{(0)} \left[ \tilde{E}^{(0)}\, V_{tt}^\pm \pm 2\,\tilde{L}_z^{(0)}\, V_{t \cf}^\pm \right]}{4\, M^{1/2}\, r^{13/4}\, \sqrt{Z_\pm}}\, .
\end{split}
\end{equation}
Here, for the ease of writing and clean presentation, we have introduced the following notations, $Y_\pm=\pm a^3 \sqrt{M} - a^2 M \sqrt{r} + a^2 r^{3/2} \pm a \sqrt{M} r^2 - 3 M r^{5/2} + r^{7/2}$, $U_{t n}^\pm= (-1)^n \left[\pm 2 a (\sqrt{M r} \mp a) \tilde{f}_{t n}^\pm + r \Delta \partial_r \tilde{f}_{t n}^\pm \right]$, and $V_{t n}^\pm = (-1)^n \left[-2 Y_\pm \tilde{f}_{t n}^\pm + r \Delta (\pm a \sqrt{M} + r^{3/2}) \partial_r \tilde{f}_{t n}^\pm \right]$, where $n = 0$ ($n=3$) is the placeholder for the coordinate $t$ ($\cf$). Moreover, the upper (lower) sign is for prograde (retrograde) motion, and $\tilde{f}_{t t}^\pm = \tilde{f}_{t t}(\th = \pi/2)$ but $\tilde{f}_{t \cf}^\pm = \pm \tilde{f}_{t \cf}(\th = \pi/2)$.

\end{document}